# KINETICS OF ELECTRODEPOSITION OF SILVER AND COPPER AT TEMPLATE SYNTHESIS OF NANOWIRES

P.G. Globa *, E.A. Zasavitsky **, V.G. Kantser **, S.P. Sidelinikova*, A.I. Dikusar*

\* Institute of Applied Physics of Academy of Sciences of Moldova,
Academiei str., 5, Chisinau, MD - 2028, Republic of Moldova
\*\* Institute of Electronic Engineering and Industrial Technologies
of Academy of Sciences of Moldova,
Academiei str., 3/3, Chisinau, MD - 2028, Republic of Moldova

**Abstract.** The results of investigation of kinetics of nanopores filling into membranes from aluminum oxide (pore diameter - 200 nm, porosity ~ 50%) at electrodeposition of copper and silver are described. It is shown, that at identical quantity of electricity passed through solutions, the degree of pores filling by metal (average thickness of a deposit) is various for copper and silver deposition. Calculated (according Faraday Law) and experimental dependences of deposition rates of these metals on quantity of electricity passed at direct and pulse currents are presented. Galvanodynamic $i$ - $v$ dependences have been obtained at various current scanning rates. The smaller rate of deposition allows to decrease concentration limitations of electrode process and to obtain higher average thickness of metal deposits and higher filling degree. The limiting values of quantity of electricity for direct and pulse currents were determined. The average thickness of silver and copper deposits was obtained. A degree of pores filling, the morphology and chemical microanalysis were studied on cross-section of the membrane, using TESCAN SEM equipped with an Oxford Instruments INCA Enerqy EDX –system.

## 1. Introduction

Artificially created nanostructures, including metal and semiconductor nanowires and nanotubes, owing to unique mechanical, magnetic, optical, thermoelectric and other properties get an increasing importance in manufacturing techniques of solid state devices used in various branches of techniques and science [1, 2].

Traditionally electrochemical methods in microelectronics are used for obtaining of thin films of various metals and semiconductors of conducting materials. Now electrodeposition is used for obtaining of nanowires. Nanowires are obtained by dimensional electrochemical deposition of metal or semiconductor perpendicular to the base surface in nano- or microcells of special patterns (templates) [1 - 7].

Aluminum oxide due to homogeneous porosity ideally suits for template synthesis of nanostructures. Besides problems of research of properties of the materials obtained by the electrodeposition method and their connection with technology of synthesis there is a complex of other problems, among them: 1) prediction of layer growth rate at electrodeposition and obtaining of required thickness for concrete time; 2) determination of nature of growth rate limitations; 3) finding of conditions of uniform electrodeposition both in the layer thickness and on the surface.

Solution of the above stated problems concerns kinetics of the concrete electrochemical process [3, 7 – 9]. The present work is intended for research of interrelation between peculiarities of kinetics of electrochemical deposition of Cu and Ag into porous $Al_2O_3$ membranes and kinetics of nanopore filling.

## 2. Experimental

Commercial aluminum oxide membranes with the thickness of 36 – 47 μm, with the pore diameter of about 200 nm and porosity of 50% were used. On one side of the membrane silver as a contact layer was put by the vacuum vapor method. On this layer deposition was made.

Electrodeposition was performed in three-electrode glass cell at room temperature. Potential was measured relative to saturated Ag/AgCl reference electrode. Electrodeposition was carried out from water solutions of electrolytes of the following composition, g/l:
1) $CuSO_4$ $5H_2O$ - 2,5; $Na_4P_2O_7$ - 100;
2) $AgNO_3$ - 40; KCNS - 300.

pH of both electrolytes was close to neutral. Cathodic current density was calculated taking into account aluminum oxide porosity. For measurement of potential and maintenance of constant current density potentiostat was used. A degree of pore filling with metal (thickness and concentration of metal) was determined by the membrane cross section, using Tescan Scanning Electron Microscope (SEM) equipped with the Oxford Instruments INCA Energy EDX system for chemical composition microanalysis. Peculiarity of the used analysis method is that the average concentration of elements in volume 2x2x2 $\mu m^3$ was fixed. Direct current density was changed within the limits of 5 - 6 $mA/cm^2$. Pulse electrodeposition by rectangular unipolar pulses of a current was carried out. The current density of the pulse $i_{avg}$ = 6 $mA/cm^2$, duration of the pulse ($\tau_p$ = 0,01s with, duration of the pause $\tau_{pp}$ =0,09 – 0,5s).

Results of SEM and EDX analysis allowed us to determine average thickness of the obtained layer and concentration of a concrete element (Ag, Cu) by depth of pores given below in the form of weight percent of the general concentration of the elements fixed in the given point of the membrane. On the basis of these data the average rate of electrodeposition $\upsilon_{avg}$ ($\mu m/hour$) and degree of pore filling with metal were calculated.

Average rate of electrodeposition was calculated according to the Faraday law:

$$V_{avg} = \frac{\mu_{avg}}{\tau} = \frac{C}{\rho}i, \qquad (1)$$

where: $\mu_{avg}$ is the average thickness of the layer, $\tau$ is the time of electrodeposition, C is the electrochemical equivalent of metal ($C_{Cu}$ = 1,18 g/A·hour and $C_{Ag}$ = 3,98 g/A hour), $\rho$ is the density of metal ($\rho_{Cu}$ = 8,96 $g/cm^3$, $\rho_{Ag}$ = 10,5 $g/cm^3$), $i$ is the current density. Calculated average rates of deposition were compared with the ones measured experimentally according to the SEM micrographs.

Galvanodynamic measurements were carried out at change of the current rate from $6 \cdot 10^3$ mA/min. to 0,6 mA/min.

### 3. Results and discussion

*Galvanostatic electrodeposition.* Fig. 1 shows SEM micrograph of the membrane cross section after electrodeposition of copper at current density 5 $mA/cm^2$ and Q = 50 $C/cm^2$.

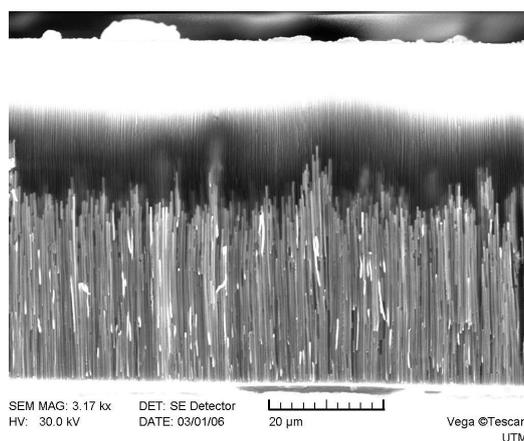

***Fig. 1***. *SEM micrograph of $Al_2O_3$ membrane cross section after Cu deposition at i = 5mA/$cm^2$, Q = 50 C/$cm^2$.*

It is seen that average thickness of the metal layer is equal to 26 $\mu m$. On the basis of the micrographs the average thickness of the electrodeposited layer was determined at deposition of both Cu and Ag at various Q. The obtained results are presented in Fig. 2.

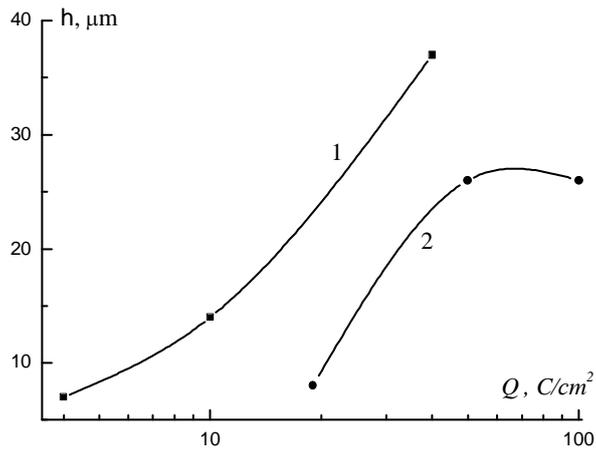

***Fig. 2.*** *Dependence of average height of metal layer h into $Al_2O_3$ membrane pores on quantity of the passed electricity at electrodeposition of Ag ($i_{Ag}$ = 6 mA/cm$^2$) (1) and of Cu ($i_{Cu}$ = 5 mA/cm$^2$) (2).*

From the results given in Fig. 2 it is seen that: a) thickness of Ag layer at equal values of Q is essentially more than that of Cu layer; b) at sufficiently high values of Q thickness of copper layer does not increase with further growth of Q, this testifies to presence of deposition rate limitations.

Higher rate of Ag deposition is caused by great values of its electrochemical equivalent. Value of the relation $(C/\rho)_{Ag} / (C/\rho)_{Cu}$ = 2,87, that is at the same values of Q the thickness of silver layer should be 3 times more than thickness of copper layer, this is qualitatively confirmed by the results shown in Fig.2.

Comparison of average rates of electrodeposition with the calculated ones (Fig. 3) shows that at Ag electrodeposition at the initial stage an essential excess of the deposition rate in comparison with the calculated one is possible (Fig. 3,a). It is due to the fact that at the initial stage deposition occurs not on the whole surface, but only on its part (probably, the most active). At increase of Q the deposition rate decreases, and at sufficiently high values of the passed charge it becomes close to the Faraday's one calculated for the whole surface of electrodeposition. This fact is important because equality of the calculated and experimentally measured rates of deposition allows us to predict obtaining of the required layer thickness.

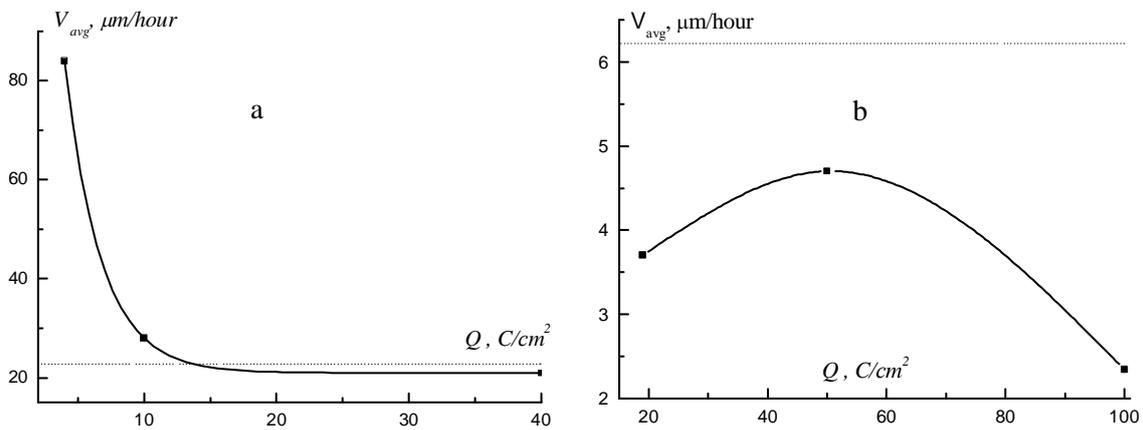

***Fig. 3.*** *Dependence of average rate of Ag (a) and Cu (b) deposition on quantity of the passed electricity. The deposition rates calculated by equation (1) are shown by the dotted line.*

It is necessary to note that for deposition of Cu layers of the same thickness as Ag, more electricity is required (Fig.3,b). However, at sufficiently high Q there can appear concentration limitations and the process rate will be determined by discharged ion diffusion to the surface. This leads to decrease in the deposition rate on greater part of the surface and, as a consequence, to lower rate of deposition in comparison with the calculated one. At sufficiently high Q, growth of the layer stops (Fig. 2, 3,b). It is take place owing to small concentration of Cu ions in the solution.

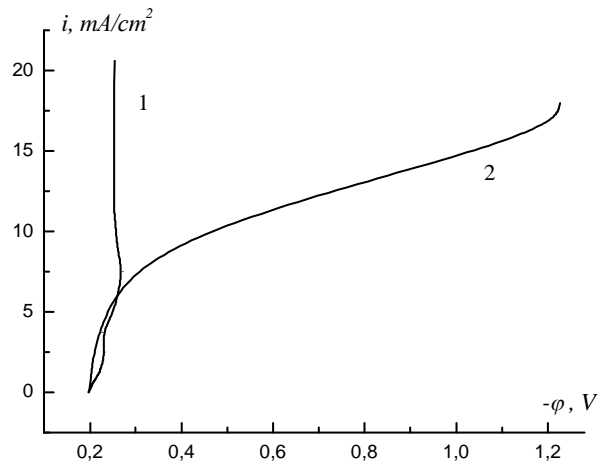

***Fig. 4.*** *Galvanodynamic polarization curves obtained at Ag electrodeposition with current change rate $6 \cdot 10^{-3}$ mA/min. (1) and 0,6 mA/min. (2).*

The opportunity of occurrence of concentration limitation follows from galvanodynamic curves (Fig. 4, measured only at Ag deposition). It is seen that the increase of current change rate leads to a sharp growth of potential and appearance of the cathodic limiting current, being a consequence of concentration limitations of the reaction rate (Fig. 4).

Presence of the concentration limitations at sufficiently high Q leads to: a) significant nonuniformity of the layer; b) appearance of dendrites on the membrane surface (Fig. 5,a). Fig.5 shows data of local microanalysis (EDX) of all the components found in a concrete point of the cross section. In connection with the above stated peculiarities of the analysis the results include not only deposited metal, but also aluminum oxide (Al, O) and the solution components.

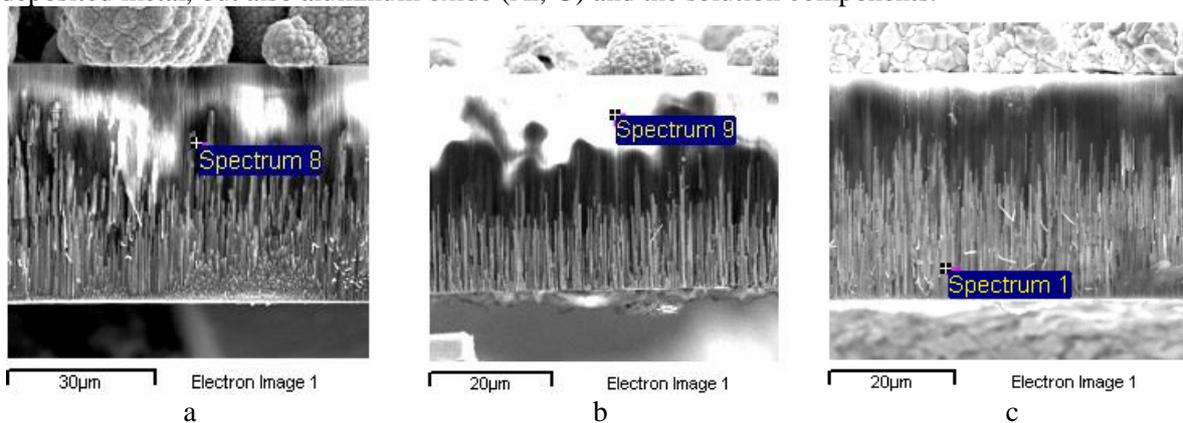

| Element | Weight% | Atomic% |
|---|---|---|
| C | 2.44 | 4.39 |
| O | 44.15 | 59.65 |
| Al | 37.35 | 29.92 |
| P | 1.60 | 1.12 |
| Cu | 14.46 | 4.92 |
| *Totals* | *100.00* | |

| Element | Weight% | Atomic% |
|---|---|---|
| O | 47.04 | 63.20 |
| Al | 45.50 | 33.51 |
| P | 2.13 | 1.48 |
| K | 2.13 | 1.17 |
| Ag | 3.20 | 0.64 |
| *Totals* | *100.00* | |

| Element | Weight% | Atomic% |
|---|---|---|
| O | 27.08 | 64.51 |
| Al | 9.19 | 12.98 |
| Ag | 63.73 | 22.52 |
| *Totals* | *100.00* | |

***Fig. 5.*** *SEM. micrographs of the membrane cross section and content of elements after Cu electrodeposition at $i = 5$ mA/cm$^2$ and $Q = 100$ C/cm$^2$ (a) and Ag at $i = 5$ mA/cm$^2$ and $Q = 10$ C/cm$^2$ (b) and at Ag electrodeposition with use of the pulse current ($i_p = 5$ mA/cm$^2$, $\tau_p = 10^{-2}$ s, $\tau_{pp} = 5 \cdot 10^{-2}$ s $Q_{avg} = 5$ C/cm$^2$) (c). Points for which distribution of elements is obtained are shown by crosses.*

Copper is found not only in the membrane pores, but also outside the layer thickness corresponding to its average value, although in smaller quantity (Fig. 5,a). Thus, separate wires can come on the surface leading to dendrite growth on the membrane surface (Fig. 5,a, the top part of the micrograph). At small Q the dendrite growth does not occur and the metal layer thickness is uniform.

Occurrence of diffusion limitations of electrodeposition rate can lead to the potential growth up to the values at which reduction of hydrogen (and as a consequence the solution alkalinizing) is possible. $Al_2O_3$ is dissolved in alkali, this can lead to breaking of isolation of separate wires. Use of similar modes seems to be impossible, this imposes restrictions on use of electrodeposition at direct current.

*Pulse electrodeposition.* Possibilities of the pulse current were investigated at Ag electrodeposition as an example. In this case the current density in the pulse $i_p$ was maintained the same as at the direct current ($i_p$ = 6 mA/cm$^2$), duration of the pulse $\tau_p$ was constant and equal to 0,01s, and duration of the pause $\tau_{pp}$ changed from 0,09 up to 0,5s. The average current density $i_{avg}$ and frequency of current pulse $f$ were calculated by equations 2 and 3:

$$i_{avg} = \frac{i_p \cdot \tau_p}{\tau_p + \tau_{pp}}. \qquad (2)$$

$$f = \frac{1}{\tau_p + \tau_{pp}}. \qquad (3)$$

Application of the pulse current allows considerable increasing of the metal layer thickness at identical quantities of the passed electricity. For example, as it follows from Fig. 2, the average thickness of Ag layer at Q = 5 C/cm$^2$ ~ 7 μm, while at the same average Q in the pulse mode it is equal to 26 μm (Fig.5,c).

As it follows from the data shown in Fig.6, the deposition rate in the pulse mode considerably exceeds the calculated one. It can be explained only by the following. When a certain potential for the pulse period during a pause is achieved it decreases slowly, therefore electroreduction during the pause time is observed. This is confirmed by the results of influence of pulse frequency on the deposition rate in comparison with the calculated ones (Fig.6). The increase in frequency means decrease of $\tau_{pp}$ (at the same value of the pulse duration (3)), and consequently greater electrodeposition rate for the period $\tau_{pp}$ (Fig.6, insert).

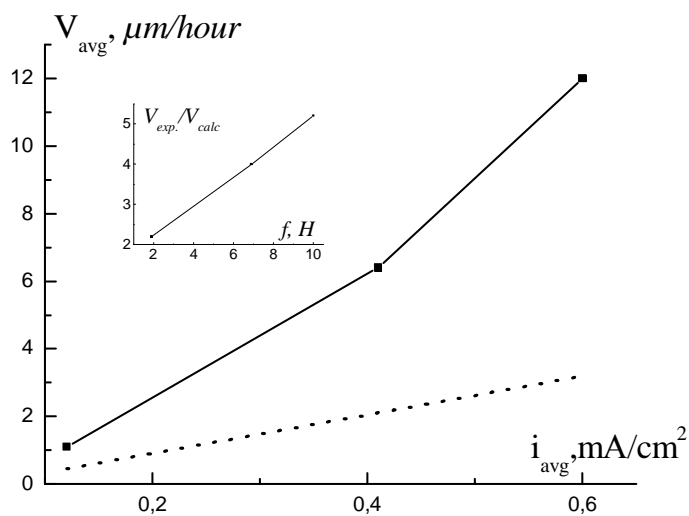

*Fig.6. Influence of average current density on rate of Ag pulse electrodeposition. Dependence calculated according to (1) is shown by the dotted line; on the insert - dependence of relation $V_{exp.}/V_{calc.}$ on frequency ($V_{exp.}$ are experimental values of the deposition rate, $V_{calc.}$ are values calculated according to (1).*

Decrease of electrodeposition average rate at pulse deposition and probability of electrodeposition during a pause (with smaller rate) allow essential increasing of degree of membrane pore filling with metal (Fig.7). Fig. 7 shows values of Ag concentration in the points located at distance h from of the pore bottom (*l* is the pore depth). It is seen that in conditions of pulse electrodeposition even at smaller Q ($Q_{avg}$) the filling degree is essentially higher.

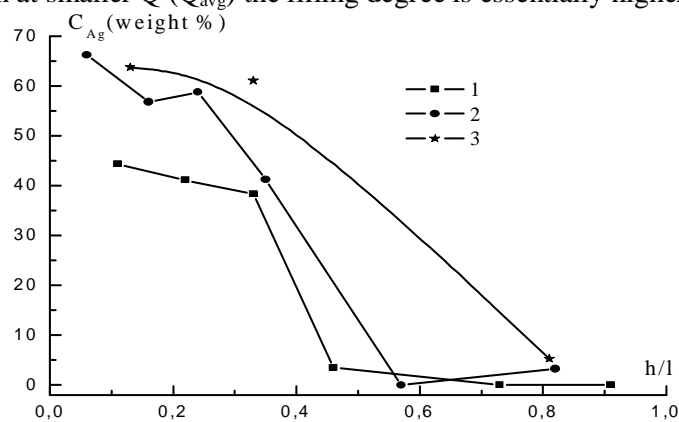

*Fig.7. Dependence of Ag concentration on dimensionless distance (h/l) (h is the distance from the pore bottom, l is the pore depth) measured at direct current deposition, i = 6 mA/cm2 (1, 2) (Q = 4 C/cm$^2$ -1, Q = 10 C/cm$^2$ - 2) and at the pulse current at $i_p$ = 6 mA/cm$^2$ and $Q_{avg.}$ = 5 C/cm$^2$ - 3.*

It seems obvious that the used modes of the pulse current are far from optimum, however the carried out experiments allow us to consider pulse electrodeposition perspective for template synthesis. Dendrites are consequence of presence of concentration limitations in any processes of electrodeposition. The present work shows their role at deposition in membrane pores with great aspect ratio A (relation of the pore depth l to the diameter d (A = l/d =180 - 235)).

It is known (e.g. [10]) that limiting cathodic currents sharply increase in pulse conditions, and therefore the probability of electrochemical obtaining of nanomaterials with high A increases too. Unfortunately, specific and weakly studied peculiarities of electrodeposition in nanopores (only a part of them being presented in this work) at present do not allow us to find optimum modes of electrodeposition, this requires further research.

## 4. Conclusions

The experimental results of kinetics of dimensional deposition in nanopores in conditions of template synthesis (Cu and Ag deposition as an example) described in the present work, show that in some cases it is possible to predict obtaining of layers of necessary thickness using galvanostatic electrodeposition (at direct current). At long electrolysis there appear concentration limitations leading under certain conditions not only to dendrite formation but also to membrane dissolution. Possibilities of pulse electrodeposition as a method for decrease of concentration limitations are shown.

The work was partially financed in the framework of RM State Program of "Nanotechnologies, new multifunctional materials and microelectronic systems".